\documentstyle[aps]{revtex}
\begin{document}
\voffset=-15mm
%
\title{Beyond the Hubbard-I Solution with a One-Pole 
\\ and a Two-Pole Self-Energy:  
Moment Approach}
\author{J J  Rodriguez - Nunez and M  Argollo de Menezes}
\address{Instituto de F\'{\i}sica, 
Universidade Federal Fluminense,\\ 
Av.\ Litor\^anea S/N, 
Boa Viagem,\\ 24210-340 Niter\'oi RJ, 
Brazil. \\e-m: jjrn@if.uff.br}
\date{\today}
\maketitle

%
%
\begin{abstract}
We have postulated a single pole 
for the self-energy, 
$\Sigma({\bf k},\omega)$, looking 
for the consequences on   
the one-particle Green function, 
$G({\bf k},\omega)$ in the 
Hubbard model.  
We find that $G({\bf k},\omega)$ satisfies 
the first two sum rules or moments of 
Nolting 
(Z. Physik {\bf 225}, 25 (1972)) for  
any values of the two unknown 
${\bf k}$ parameters of  
$\Sigma({\bf k},\omega)$. In order 
to find these two 
parameters we have used the third 
and four sum rules of 
Nolting. $G({\bf k},\omega)$ turns 
out to be identical to the 
one of Nolting (Z. Physik {\bf 225}, 
25 (1972)), which is 
beyond a Hubbard-I solution. Furthermore, 
we have also postulated a 
two-pole Ansatz for $\Sigma({\bf k},\omega)$ 
which is equivalent to a three-pole 
Ansatz for $G({\bf k},\omega)$. In 
the latter case, we 
have fixed two of the frequencies. This 
Green's function is also beyond the 
Hubbard-I solution. We  
present numerical results for the band structure and 
the spectral weights. 
\\
Pacs numbers: 74.20.-Fg, 74.10.-z, 
74.60.-w, 74.72.-h
\end{abstract}

%
%
%
%

	After the discovery of 
the high-$T_c$ materials
\cite{Bednorz-Muller}, the study 
of correlations 
has gained interested due to the 
fact that there is 
the belief\cite{Anderson} 
that the normal properties 
of these materials 
could be explained in the 
framework of the Hubbard 
model\cite{HubbardI}, since 
electron correlations 
are strong, i.e., the on-site 
electron-electron 
repulsions $U$ are much 
larger than the energies 
associated with the hybridization 
of atomic orbitals 
belonging to different 
atoms\cite{Fulde}. 
This Hamiltonian is a kind 
of minimum model\cite{tasaki} which 
takes into account quantum 
mechanical motion of 
electrons in a solid, 
and nonlinear repulsion 
between electrons. Even 
though this 
model is too simple to 
describe solids faithfully, 
serious 
theoretical studies have 
revealed that to understand its 
various properties is a 
very difficult task, since 
is the simplest many-body Hamiltonian 
one can write down and which cannot be 
reduced to a single-particle 
theory\cite{Assa}. Its study will prove 
useful in developing 
various notions and techniques 
in statistical physics of 
many degrees of freedom. Besides 
cuprate superconductors, the 
Hubbard model may also be applicable to 
describing the metal - 
insulating transition in materials like 
$La_xSr_{1-x}TiO_3$ and 
$V_{2-y}O_3$ in which paramagnetic 
metal, antiferromagnetic 
insulator, and a phase 
antiferromagnetic 
metal are clearly 
separated in temperature - 
pressure phase 
diagram\cite{all}. All 
these features, found in the 
Hubbard model, 
make it suitable to 
reproduce experimental data. 
Additionally, 
the study of correlations by 
itself in the Hubbard model 
is a rewarding 
task since it will shed light 
on still unsolved points of 
the novel materials. 
For example, at high 
temperatures ($T_c~30-130~K$) these 
HTSC cuprates, which are 
poor conductors, become superconductors. 
This feature is strange 
indeed because the Coulomb 
repulsion is 
strong. Furthermore, the 
behavior of these materials 
at $T~>~T_c$ 
is even more puzzling than 
the superconductivity itself. 
Contrary to the 
predictions of the Fermi 
liquid theory, the resistivity at 
$T~>~T_c$ and optimum doping is 
linear in temperature, i.e., 
$R~\approx~T$. 
This suggests a very strong 
scattering of elementary 
excitations. 
A discussion of the possible breakdown 
of Fermi liquid theory is given 
in Ref.\cite{Alexei}. 

	In this paper, we will adopt  
a given pole structure in the self-energy 
and see its effect 
on the one-particle Green 
function. This is done 
with the idea of 
reducing the amount of 
computational effort which 
is implicit in 
the moment approach of Nolting 
which is normally constrained to two 
poles in $G({\bf k},\omega)$. At the 
same time, without additional effort, 
we monitor the self-energy which is 
an important quantity for life-time 
considerations. 
We find that by using this 
approach (one pole in 
$\Sigma({\bf k},\omega)$ 
and the sum rules for the spectral 
functions), the one-particle 
Green function goes beyond the 
Hubbard-I solution, i.e., it 
identically satisfies the 
first two sum rules 
and its parameters have 
to be calculated 
self-consistently. This 
has motivated us to 
consider two poles 
in the self-energy 
and to see the effects on 
$G({\bf k},\omega)$. 
The general result 
that we gain is that 
$N$-poles in 
$\Sigma({\bf k},\omega)$ 
is equivalent to 
$N+1$-poles in $G({\bf k},\omega)$. 
Along the way, we discuss the 
possibility of 
obtaining a third peak 
in $G({\bf k},\omega)$   
by means of this technique. 

    	The model we study is the  
Hubbard model\cite{HubbardI}
\begin{eqnarray}\label{Ham}
H = t_{\vec{i},\vec{j}}
c_{\vec{i}\sigma}^{\dagger}
c_{\vec{j}\sigma}
   + \frac{U}{2} n_{\vec{i}\sigma}
n_{\vec{i}\bar{\sigma}}   
   - \mu c^{\dagger}_{\vec{i}\sigma}
c_{\vec{i}\sigma}~~,
\end{eqnarray}
where $c_{\vec{i}\sigma}^{\dagger}
$($c_{\vec{i}\sigma}$) are creation
(annihilation) electron operators with 
spin $\sigma$. $n_{\vec{i}
\sigma} \equiv 
c_{\vec{i}\sigma}^{\dagger}
c_{\vec{i}\sigma}$. 
$U$ is the local interaction, 
$\mu$ the chemical
potential and we 
work in the grand 
canonical ensemble. We have 
adopted Einstein 
convention for repeated 
indices, i.e., for the 
$N_s$ sites $\vec{i}$, the 
$z$ nearest-neighbor sites 
and for spin up and down ($\sigma 
= \pm 1$). 
$t_{\vec{i},\vec{j}} = 
-t$, for n.n. 
and zero otherwise.

	The one-particle 
Green function, 
$G({\bf k},\omega)$, is 
expressed in terms of 
the self-energy, 
$\Sigma({\bf k},\omega)$, as
\begin{equation}\label{1PGF}
G({\bf k};\omega) \equiv  
\frac{1}{\omega - \varepsilon_{\bf k} - 
\Sigma({\bf k},\omega)} ~~~,
\end{equation}
\noindent where 
$\epsilon({\bf k}) =
-2 t (cos(k_xa) + cos(k_ya))$, 
and $\varepsilon_{\bf k} = 
\epsilon({\bf k}) - \mu$. 

	We adopt the following Ansatz for 
$\Sigma({\bf k},\omega)$:
\begin{equation}\label{ansatz}
\Sigma({\bf k},\omega) \equiv \rho U +  
\frac{\bar{\alpha}({\bf k})}{\omega - 
\Omega_{\bf k}} ~~~.
\end{equation} 

	As $\Sigma({\bf k},\omega)$ 
has dimensions of energy, 
the still unknown parameter 
$\bar{\alpha}_{\bf k}$ has dimensions of 
(energy)$^2$. $\bar{\alpha}({\bf k})$ is ]
kind of a spectral weight and 
$\Omega_{\bf k}$ is the energy spectrum 
of the self-energy. We will 
calculate $\bar{\alpha}({\bf k})$ and 
$\Omega_{\bf k}$. Along the 
way, we will give a simple physical 
interpretation for 
$\bar{\alpha}({\bf k})$. We have included 
the Hartree shift directly 
in the self-energy.  
In Eq. (\ref{ansatz}), $\rho$ is 
the carrier concentration per spin orientation, i.e., 
$\rho = n/2$. We are 
assuming that we are in the 
paramagnetic phase\cite{magnetic}.

	By using Eq. (\ref{ansatz}) into 
Eq. (\ref{1PGF}), we get 
that the one-particle Green function has two poles. It can be 
written as
\begin{equation}\label{twopoles}
G({\bf k},\omega) = \frac{\alpha_1({\bf k})}{\omega - 
\omega_1({\bf k})} + 
\frac{\alpha_2({\bf k})}{\omega - \omega_2({\bf k})} ~~~,
\end{equation}
\noindent where 
\begin{eqnarray}\label{solvin}
\omega_1({\bf k}) &=& \frac{1}{2} 
\left[ \Omega_{\bf k} + \xi_{\bf k} + 
\left[\left(\Omega_{\bf k} - 
\xi_{\bf k} \right)^2 + 
4\bar{\alpha}({\bf k}) 
\right] ^{1/2} \right] ~~~, 
\nonumber \\
\omega_2({\bf k}) &=& \frac{1}{2} 
\left[ \Omega_{\bf k} + \xi_{\bf k} - 
\left[\left(\Omega_{\bf k} - 
\xi_{\bf k} \right)^2 + 
4\bar{\alpha}({\bf k}) 
\right]^{1/2} \right] ~~~, 
\nonumber \\ 
\alpha_1({\bf k}) &=& 
\frac{\omega_1({\bf k}) - 
\Omega_{\bf k}}
{\omega_1({\bf k}) - 
\omega_2({\bf k})} 
~~~, \nonumber \\
\alpha_2({\bf k}) &=& 
\frac{\omega_2({\bf k}) - 
\Omega_{\bf k}}
{\omega_2({\bf k}) - 
\omega_1({\bf k})} ~~~ .
\end{eqnarray}

	From Eqs. (\ref{solvin}) 
we inmediately see that
the following sum rules or 
moments\cite{Nolting} are identically 
satisfied:
\begin{eqnarray}\label{firstwo}
\alpha_1({\bf k}) + \alpha_2({\bf k}) = 
1 ~~~ &,& \nonumber \\
\alpha_1({\bf k}) \omega_1({\bf k}) + 
\alpha_2({\bf k}) \omega_2({\bf k}) = 
\xi_{\bf k} ~~~ &,& \nonumber \\
\xi_{\bf k} = \epsilon({\bf k}) + 
\rho U ~~~ &.&
\end{eqnarray}

	Eqs. (\ref{firstwo}) are the 
first two sum rules for the 
spectral functions of Nolting\cite{Nolting}. 
In order to evaluate 
$\bar{\alpha}({\bf k})$ and $\Omega_{\bf k}$, 
we use the next two sum rules of 
Nolting. This gives:
\begin{eqnarray}\label{nextwo}
\omega_1^2({\bf k}) 
\alpha_1({\bf k}) + 
\omega_2^2({\bf k}) 
\alpha_2({\bf k}) = 
\xi^2_{\bf k} + 
\bar{\alpha}({\bf k}) &=&  
a_2({\bf k}) ~~~ , \nonumber \\
\omega_1^3({\bf k}) 
\alpha_1({\bf k}) + 
\omega_2^3({\bf k}) 
\alpha_2({\bf k}) = \xi^3_{\bf k} + 
(\Omega_{\bf k} + 2\xi_{\bf k})
\bar{\alpha}({\bf k}) 
&=& a_3({\bf k}) ~~~ ,
\end{eqnarray}
where $a_2({\bf k})$,  
$a_3({\bf k})$ are given in 
Ref.\cite{Nolting}-\cite{Micnas-et-al} 
as
\begin{eqnarray}\label{a2-a3}
a_2({\bf k}) = 
\epsilon^2({\bf k}) + 
2\rho U \epsilon({\bf  k}) + 
\rho U^2 ~~~ &,& \nonumber \\
a_3({\bf k}) = 
\epsilon^3({\bf k}) + 
3 U \epsilon^2({\bf k}) 
(2+\rho) \rho U^2 
\epsilon({\bf k}) + 
\rho (1-\rho) U 
B({\bf k}) + 
\rho U^3 ~~~ &,& 
\end{eqnarray}
\noindent
where $B({\bf k})$, in the spherical 
approximation of Nolting, is 
given by
\begin{equation}\label{B}
\rho (1-\rho)B({\bf k}) =  
\sum_{j = 1}^2 \sum_{\bf k} 
\left[ \frac{2}{U} 
\left( \omega_j({\bf k}) - 
\epsilon({\bf k}) 
\right) -1 \right] 
\alpha_j({\bf k}) 
f(\omega_j({\bf k})) 
\epsilon({\bf k}) 
~~~ , 
\end{equation}
\noindent where $f(x)$ is the Fermi 
distribution function. 
By solving 
Eqs. (\ref{nextwo}) we find:
\begin{eqnarray}\label{solutions}
\bar{\alpha}({\bf k}) = \rho (1 - \rho ) 
U^2 ~~~ &,& \nonumber \\
\Omega_{\bf k} = (1 - 
\rho) U + 
B({\bf k}) ~~~ &,&
\end{eqnarray}
\noindent $B({\bf k})$ is the 
narrowing band parameter 
defined by Nolting. This 
narrowing band parameter has to be 
calculated 
self-consistently and 
in the spherical 
approximation of 
Nolting is ${\bf k}$ 
independent. By 
combining Eqs. 
(\ref{solvin},\ref{solutions}) 
we find
\begin{equation}\label{cloNol}
\omega_1({\bf k}) = 
\frac{1}{2}\left[H({\bf k}) + 
[K({\bf k})]^{1/2}\right] ~~~ ; ~~~ 
\omega_2({\bf k}) = 
\frac{1}{2}\left[H({\bf k}) - 
[K({\bf k})]^{1/2}\right] ~~~ , 
\end{equation}
\noindent where
\begin{eqnarray}\label{Noltingregained}
H({\bf k}) \equiv 
\epsilon({\bf k})  
U + B({\bf k}) 
~~~ &,&
 \nonumber \\
K({\bf k}) \equiv 
\left(\epsilon({\bf k}) - U - 
B({\bf k}) \right)^2 + 4 \rho U 
\left(\epsilon({\bf k})  
- B({\bf k}) \right) ~~~ &.&
\end{eqnarray}
	Eqs. (\ref{Noltingregained}) 
are nothing that the 
solutions given by 
Nolting\cite{Nolting} by solving  the 
four unknown 
$\alpha_i({\bf k})$ 
and $\omega_i({\bf k})$, 
with $i = 1,2$.  
We have regained Nolting's solutions 
in a much easier way, starting 
from the self-energy, while 
Nolting does it from the 
Green function itself. At the 
same time we have been able to 
identify the 
self-energy for the 
two-pole ansatz 
for the one-particle 
Green function. 
Even more, 
we see that the 
self-energy has 
the meaning of being an 
expansion in powers 
of $U$. In our case, 
due to the choice of 
a single pole for the self-energy, 
$\bar{\alpha}({\bf k})$ is of second 
order in $U$. Remember that the 
first order expansion has been 
taken into account by 
the Hartree shift. Then, if  
we took two poles for 
the self-energy, 
then $\bar{\alpha}({\bf k})$ has to 
go up to order $U^3$. 
This is a 
conjeture we will justify  
later. We would like to 
point out that in the 
atomic limit, i.e., $t = 0$, we recover 
the two Hubbard bands, i.e., 
\begin{eqnarray}\label{atomiclimit}
\omega_1({\bf k}) = - 
\mu + U ~~~ &;& ~~ 
\omega_2({\bf k}) = - 
~~~ \mu ~~~, \nonumber \\
\alpha_1({\bf k}) = \rho ~~~ &;& ~~ 
\alpha_2({\bf k}) = 
1 - \rho ~~~ ,
\end{eqnarray}
\noindent as it 
should be. As our 
solutions given by 
Eqs. (\ref{Noltingregained}) 
satisfy the first four sum rules or 
moments, our Green function 
is beyond the Hubbard-I 
solution\cite{Izyumov-Sckryabin} 
and as a consequence we have an 
improved solution\cite{Micnas-et-al}. 
The drawback of the Hubbard-I solution, 
i.e., a gap for any value of the 
interaction, was pointed out 
by Laura 
Roth\cite{LRot} many years 
ago. This gap exists, for 
all values of $U$, in 
the spherical 
approximation of 
Nolting\cite{Nolting}-
\cite{Micnas-et-al}. 
It is known that the Hubbard-I 
solution does not 
show ferromagnetism. 
This difficulty is solved 
by the presence of the 
$B$-term in Nolting 
solution which is spin-dependent. 

	Eqs. (\ref{solvin}) contain 
interesting solutions. One of them 
we consider next. For 
example, if we take 
\begin{equation}\label{halff}
\omega_1({\bf k})~\equiv~
-~\omega_2({\bf k}) ~~~ ,
\end{equation}
\noindent we inmediately get
\begin{eqnarray}\label{omegas}
\Omega_{\bf k} &=& \xi_{\bf k} 
~~~~ , \nonumber \\
\omega_1({\bf k}) = 
\left[ \xi^2_{\bf k} + 
\rho (1-\rho) U^2 \right]^{1/2}  
&\equiv& E_{\bf k} ~~~ ,
\end{eqnarray}
\noindent and 
\begin{eqnarray}\label{alphas}
\alpha_1({\bf k}) 
&=& \frac{1}{2} \left( 1 
+ \frac{\xi_{\bf k}}
{E_{\bf k}} \right) ~~~ , 
\nonumber  \\
\alpha_2({\bf k}) = 
1 - \alpha_1({\bf k}) &=&  
\frac{1}{2} \left( 1 
- \frac{\xi_{\bf k}}
{E_{\bf k}} \right)  ~~~ .
\end{eqnarray}

	Eqs. 
(\ref{omegas}-\ref{alphas}) 
are the well-known 
spin-density-wave (SDW) solutions at 
half-filling, with a  
gap given by $U/2 -
(-U/2) = U$\cite{Scalapino}. 
Thus, our solutions 
reduce to known results 
for some values 
of parameter space. 
Chubukov and Morr\cite{CM} 
have discussed 
the mean-field SDW result 
within the effective 
spin-fermion model. 
According to these authors a one-pole 
self-energy close to 
half-filling includes 
only exchange of longitudinal 
fluctuations. Substituting this diagram 
(self-energy) into 
Dyson equation 
produces inmediately 
the mean-field 
SDW result. Chubukov 
and Morr's 
analysis\cite{CM} is 
based on the theory 
of Kamp and Schrieffer\cite{KS} where 
it is assumed that 
the longitudinal spin 
susceptibility, 
$\chi_s({\bf q},\omega)$ has 
a $\delta$-functional 
peak at zero frequency and momentum 
transfer $Q = (\pi,\pi)$, i.e., 
$\chi_s({\bf q},\omega) = \frac{1}{4}
\delta({\bf q}-Q)\delta(\omega)$. 
Chubukov 
and Morr's paper studies the 
Fermi surface evolution 
with interaction. 
They find that the SDW 
mean-field 
solution (valid for 
large U) yields a small 
Fermi surface centered at 
$(\pi/2,\pi/2)$. 
Notice, however that they include 
a term $t'$ in the 
free-energy band. Let's 
mention that the SDW 
state has not been observed in FLEX 
calculations\cite{Bennemann} where 
short-wavelength spin fluctuations are 
taken into account.  
By using the Green 
function given by Eqs. 
(\ref{1PGF},\ref{Noltingregained}), 
we can evaluate 
the dynamical conductance, 
$\sigma(\omega)$, vs $\omega$, 
away from half-filling\cite{Scalapino}. 
We mention that 
the two-pole one-particle 
Green function 
(Eq. (\ref{1PGF}) has been 
used to calcultate 
the static spin 
susceptibility, 
$\chi_s(T)$,
\cite{Nolting}-\cite{Micnas-et-al} 
and the results 
compare rather well with Quantum 
Monte Carlo simulations\cite{Singer}. 
Another quantity which 
can be easily calculated is the 
specific heat, $c_V$, vs temperature, 
$T$\cite{Micnas-et-al}. 
See also Ref.\cite{Elk}. We would like to 
comment on the fact that our two energy 
branches (second line of Eq. (\ref{omegas})) 
produce a gap with a value of $U$. This 
gap is evidently no perturbative contrary 
to the view of Ref.\cite{KS}.

	Now, let us propose a two-pole 
Ansatz for the self-energy, 
given by
\begin{equation}\label{twopolesel}
\Sigma({\bf k},\omega) \equiv \rho U + 
\frac{\bar{\alpha}({\bf k})}
{(\omega - \Omega_1({\bf k}))(\omega - 
\Omega_2({\bf k}))} ~~~ .
\end{equation}

	The assumption of Eq. 
(\ref{twopolesel}) is 
equivalent to have 
a three-pole structure 
for the one-particle 
Green function, i.e.,
\begin{equation}\label{threepolegreen}
G({\bf k},\omega) \equiv 
\frac{\alpha_1({\bf k})}
{\omega - 
\omega_1({\bf k})}  + 
\frac{\alpha_2({\bf k})}
{\omega - \omega_2({\bf k})} + 
\frac{\alpha_3({\bf k})}{\omega - 
\omega_3({\bf k})} ~~~ .
\end{equation}

	The $\omega_j({\bf k})$'s, 
j=1,2,3, 
are the roots of the following 
cubic equation
\begin{equation}\label{cubic}
\omega^3 - A_1({\bf k}) \omega^2 + 
A_2({\bf k}) \omega - 
A_3({\bf k}) = 0 
~~~ ,
\end{equation}
\noindent where
\begin{eqnarray}\label{A-s}
A_1({\bf k}) \equiv \Omega_1({\bf k}) + 
\Omega_2({\bf k}) + \xi_{\bf k} 
~~~ ; ~~~ A_2({\bf k}) 
\equiv \xi_{\bf k}
\left( \Omega_1({\bf k}) + 
\Omega_2({\bf k}) \right) 
 + \Omega_1({\bf k}) \Omega_2({\bf k}) 
~~~ &,& \nonumber \\  
A_3({\bf k}) \equiv 
\bar{\alpha}({\bf k}) +  
\Omega_1({\bf k}) 
\Omega_2({\bf k}) 
\xi_{\bf k} ~~~ &.&
\end{eqnarray}

	The $\alpha_j({\bf k})$'s, 
j=1,2,3, 
are given by
\begin{eqnarray}\label{alphas-3}
\alpha_1({\bf k}) \equiv ] 
\frac{(\omega_1({\bf k}) - 
\Omega_1({\bf k})) 
(\omega_1({\bf k}) - 
\Omega_2({\bf k}))}
{(\omega_1({\bf k}) - 
\omega_2({\bf k})) 
(\omega_1({\bf k}) - 
\omega_3({\bf k}))} ~~~ 
&,& \nonumber \\
\alpha_2({\bf k}) \equiv 
\frac{(\omega_2({\bf k}) - 
\Omega_1({\bf k})) 
(\omega_2({\bf k}) - 
\Omega_2({\bf k}))}
{(\omega_2({\bf k}) - 
\omega_1({\bf k})) 
(\omega_2({\bf k}) - 
\omega_3({\bf k}))} ~~~ &,& 
\nonumber \\
\alpha_3({\bf k}) \equiv 
\frac{(\omega_3({\bf k}) - 
\Omega_1({\bf k})) 
(\omega_3({\bf k}) - 
\Omega_2({\bf k}))}
{(\omega_3({\bf k}) - 
\omega_1({\bf k})) 
(\omega_3({\bf k}) - 
\omega_2({\bf k}))} ~~~ &.& 
\end{eqnarray}

	Now, as we are after 
a simple 
solution to our Eqs. 
(\ref{cubic}-\ref{alphas-3}), 
we are going to 
impose the strong condition on the cubic equation 
(Eq. (\ref{cubic})) that 
\begin{equation}\label{omega0}
\omega_3({\bf k}) = 0 ~~~ .
\end{equation}

	Then, Eq. (\ref{cubic}) becomes
\begin{eqnarray}\label{cuadratic}
\omega^2 - A_1({\bf k}) \omega + 
A_2({\bf k}) = 0 ~~~ &,& \nonumber \\
\bar{\alpha}({\bf k}) = - ~\Omega_1({\bf k}) 
\Omega_2({\bf k}) \xi_{\bf k} ~~~ 
&.& 
\end{eqnarray}

	If we  
require that the other two roots are antisymmetric, we get
\begin{eqnarray}\label{kondopeak}
A_1({\bf k}) = \Omega_1({\bf k}) + 
\Omega_2({\bf k}) + \xi_{\bf k} = 0 ~~~ 
&,& \nonumber \\
\Omega_{\bf k}^2 = - 
\left[ \Omega_1({\bf k}) 
\Omega_2({\bf k})) + 
\xi_{\bf k} \left( \Omega_1({\bf k}) + 
\Omega_2({\bf k}) \right) \right] ~~~ 
&.&
\end{eqnarray}
We would like to point out that this 
condition (the first line of Eq. 
(\ref{kondopeak})) 
is not really necessary. It 
has been used here to 
keep the algebra as simple as 
possible. Otherwise, higher order 
moments will be needed.

	It is easy to express the 
$\alpha_j({\bf k})$'s as
\begin{eqnarray}\label{alphastepkondo}
\alpha_1({\bf k}) &=& \frac{1}
{2\Omega^2_{\bf k}} \left( 
\Omega_{\bf k} - \Omega_1({\bf k}) 
\right) \left( 
\Omega_{\bf k} - \Omega_2({\bf k}) \right) 
~~~ , \nonumber \\
\alpha_2({\bf k}) &=& \frac{1}
{2\Omega^2_{\bf k}} \left( 
\Omega_{\bf k} + 
\Omega_1({\bf k}) 
\right) \left( 
\Omega_{\bf k} + \Omega_2({\bf k}) 
\right) ~~~ , \nonumber \\
\alpha_3({\bf k}) &=& - 
\frac{\Omega_1({\bf k}) \Omega_2({\bf k})} 
{\Omega_{\bf k}^2} ~~~ .
\end{eqnarray}
	
	The first two sum rules 
or moments are satisfied 
identically since
\begin{eqnarray}\label{firstwo1}
\alpha_1({\bf k}) + \alpha_2({\bf k}) + 
\alpha_3({\bf k})  = 1 ~~~ &,& 
\nonumber \\
\Omega_{\bf k} \left( \alpha_1({\bf k}) - 
\alpha_2({\bf k}) \right) = 
- \left( \Omega_1({\bf k}) + 
\Omega_2({\bf k}) \right) = 
\xi_{\bf k} ~~~ &.&
\end{eqnarray}
	The second moment is 
satisfied due to the fact that 
$A_1({\bf k}) = 0$. 
(See Eq. (\ref{kondopeak})). 
Let us point out that the 
first sum rule is always satisfied, 
without the requirement that 
$\omega_3({\bf k}) = 0$. 

	The third and four 
moments are given by
\begin{eqnarray}\label{twoothersums}
\Omega_{\bf k}^2 
\left( \alpha_1({\bf k}) +  
\alpha_2({\bf k}) \right) 
= \Omega_{\bf k}^2 + 
\Omega_1({\bf k}) 
\Omega_2({\bf k}) = 
a_2({\bf k}) ~~~ 
&,& \nonumber \\
\Omega_{\bf k}^3\left( 
\alpha_1({\bf k}) -  
\alpha_2({\bf k}) \right) 
= -~ \Omega_{\bf k}^2 
\left(  \Omega_1({\bf k}) + 
\Omega_2({\bf k}) \right) = 
a_3({\bf k}) ~~~ &.&
\end{eqnarray}
	
	Combining Eqs. 
(\ref{firstwo1},\ref{twoothersums}) 
we end with 
\begin{equation}\label{omega2}
\Omega_{\bf k}^2~=
~\frac{a_3({\bf k})}
{a_1({\bf k})} ~~~ .
\end{equation}

	The other unknowns 
are easily found. 
They are:
\begin{eqnarray}\label{final}
\Omega_{1,2}({\bf k}) = 
\frac{- \xi_{\bf k} 
\pm \left[ \xi_{\bf k}^2 + 
4(\Omega_{\bf k}^2 - 
a_2({\bf k}) ) \right]^{1/2}}
{2} ~~~ &,& \nonumber \\
\alpha_{1,2}({\bf k}) = \frac{\xi_{\bf k} 
\left( a_2({\bf k}) 
\pm \left[ \xi_{\bf k} a_3({\bf k}) 
\right]^{1/2} \right) }{2 a_3({\bf k})} 
~~~ &,& \nonumber \\
\alpha_3({\bf k}) = 1 - 
\frac{\xi_{\bf k} a_2({\bf k})}
{a_3({\bf k})} ~~~ &,& \nonumber \\
\bar{\alpha}({\bf k}) = 
a_3({\bf k}) - \xi_{\bf k} 
a_2({\bf k}) ~~~ &.& 
\end{eqnarray}

	Now we are in a position 
to justify the 
fact that $\Sigma({\bf k},\omega)$ 
represents an 
expansion in powers of $U$. 
(See the discussion 
after Eq. (\ref{Noltingregained})). From  
the four line of Eq. (\ref{final}) we can 
calculate the explicit form of the 
self-energy weight. 
We use the expressions 
for $a_2({\bf k})$ and $a_3({\bf k})$. 
The result is 
\begin{equation}\label{self-energyweight}
\bar{\alpha}({\bf k}) = \rho (1-\rho)U^2 
\left[ U + 2\epsilon({\bf k}) + 
B({\bf k})  \right]  ~~~ .
\end{equation}
Then, we see from Eq. 
(\ref{self-energyweight}) 
that we have an expansion 
in terms of $U$. Just 
imagine the atomic limit!. 
Let us say 
that the solutions we have found 
(Eqs. (\ref{threepolegreen},
\ref{omega0},
\ref{omega2}-\ref{self-energyweight})) 
can also be obtained from 
the three pole Ansatz 
for the one-particle 
Green function, 
if we require that the 
first four moments are 
satisfied together with  
the conditions of a peak at 
zero frequency and the other two peaks 
at antisymmetric frequencies.   
In Eq. 
(\ref{self-energyweight}) 
$B({\bf k})$ is given by
\begin{equation}\label{B1}
\rho (1-\rho)B({\bf k}) =  
\sum_{j = 1}^3 \sum_{\bf k} 
\left[ \frac{2}{U} 
\left( \omega_j({\bf k}) - 
\varepsilon_{\bf k} 
\right) -1 \right] \alpha_j({\bf k}) 
f(\omega_j({\bf k})) \epsilon({\bf k}) 
~~~ . 
\end{equation}

	In Fig. 1 we present 
the energy spectra, 
i.e., $\omega_i(\vec{k})$, 
$i=1,2,3$ vs 
$\vec{k}$ along the 
diagonal of the Brillouin zone, i.e., 
$k_x = k_y = k$ for $T/t = 0.5$, 
$\rho = 0.5$ and 
$\mu = U/2$, with $U/t = 7.0$. 
In Fig. 2 we show 
the spectral weights, $\alpha_i({\bf k})$, 
$i = 1,2,3$, for ${\bf k}$ along the diagonal 
of the Brillouin zone for the same parameters 
as in Fig. 1. In Fig. 3 we plot $\Omega_i({\bf k})$, 
$i = 1,2$ along the diagonal of the Brillouin zone 
for the same parameters as in the first two figures. 
We observe from all these figures that there are 
regions of ${\bf k}$ which are forbidden, i.e., for 
which $\omega_1({\bf k}$ become imaginary. We have 
neglected these points. However, they could be taken 
into account if we include lifetime effects. Our 
choice of a delta function for the spectral 
function, $A({\bf k},\omega)$, forbids us to consider 
imaginary frequencies.

	In summary we have proposed a 
self-energy of $N$ poles which 
implies that the 
Green function is composed of $N+1$ poles. 
This has been accomplished by the use of 
Dyson's equation (Eq. \ref{1PGF}). 
For the case of a one-pole for the 
self-energy we have 
reproduced with minor effort 
the Nolting's 
solution\cite{Nolting}-\cite{Micnas-et-al}. 
The one-pole 
self-energy leading to Nolting's 
solution puts on 
firm grounds the results found 
in Refs.\cite{Micnas-et-al} 
where it was argued that the two 
branches of the one-particle 
Green function were due to a 
single branch in the 
self-energy. For the 
case of two-pole Ansatz for the 
self-energy we have worked out the 
simple case of what we call 
the Kondo peak. In the later 
case, we have 
been able to solve all the unknown 
parameters. In 
both cases, the first 
two sum rules are 
satisfied identically. 
In both cases, 
our Green's function 
goes beyond the Hubbard-I 
solution because 
we have imposed the condition 
that the third and 
four moments be satisfied.  
We would like to 
say that we have 
found the Green's function for a 
strongly correlated system, keeping 
the full $({\bf k},\omega)$ - 
dependence. With our rather simple 
approach we have been able to have a 
Green's function with 
either two poles or three 
poles. Our approch 
is somewhat different from the one of 
the limit of infinite 
dimensions\cite{Vollhardt} 
where only 
the dynamical properties 
are taken into 
account leaving aside the study 
of the long range 
behavior. The moment 
approach is a 
reliable tool to study 
strongly correlated 
electronic systems, 
in particular, the 
Hubbard model. A 
recent calculation 
by Nolting, Jaya and 
Rex\cite{NJR} has 
applied it to the 
periodic Anderson 
model, where the 
relevant quantity of study is the 
self-energy.  
\begin{center}
{\Large Acknowledments}
\end{center}
We would like to thank
the CNPq (project No.
300705/95-6)
and also from CONICIT 
(project F-139).
We thank Mar\'{\i}a 
Dolores Garc\'{\i}a
for reading the manuscript. Fruitful 
discussions with Prof. M.A. 
Continentino, Prof. E. Anda, Prof. M.S. Figueira, 
Dr. M.H. Pedersen and 
Prof. H. Beck are fully 
appreciated.\\

\vspace{1.6cm}

\begin{center}
{\Large Figures}\\
\end{center}
\vspace{0.8cm}

\noindent Fig. 1. The self-consistent energy 
spectra, $\omega_i(\vec{k})$, $i=1,2,3$, as 
function of momentum along the diagonal of the 
Brillouin zone for $U/t = 7.0$, $\rho = 0.5$, 
$T/t = 0.5$.\\
\vspace{0.3cm}

\noindent Fig. 2. $\alpha_i({\bf k})$ vs ${\bf k}$, 
$i = 1,2,3$, with the same parameter of Fig. 1.  

\vspace{0.3cm}

\noindent Fig. 3. $\Omega_i({\bf k})$ vs ${\bf k}$, 
$i = 1,2$  
for the same parameters of Fig. 1. 

\begin{references}
\bibitem{Bednorz-Muller}
      	J. Bednorz and K.A. M\"uller,
      	Z. Phys. B~{\bf 64}, 189 (1986).
\bibitem{Anderson}
	P.W. Anderson, Science 
	{\bf 235}, 1196 (1987); 
	{\it Frontiers and Borderlines 
	in Many Particle Physics}. 
	North Holland, Amsterdam (1988).
\bibitem{HubbardI}
 	J. Hubbard, Proc. R. Soc. 
	London A {\bf 276}, 238 (1963); 
	{\bf 28}, 401 (1964); M. 
	Gutzwiller, Phys. Rev. Lett. 
	{\bf 10}, 159 (1963); 
	J. Kanamori, 
	Prog. Theor. Phys. 
	{\bf 30}, 275 (1963).
\bibitem{Fulde}
	Peter Fulde, {\it Electron 
	Correlations in Molecules 
	and Solids}. Springer-Verlag 
	(1993). 2nd Edition.
\bibitem{tasaki}
	Hal Tasaki, {\it 
	The Hubbard Model: 
	Introduction and 
	Some Rigorous Results}. 
	Cond-Matt/9512169.
\bibitem{Assa}
	Assa Auerbach, {\it 
	Interacting Electron and 
	Quantum Magnetism}. 
	Springer-Verlag (1994). 
	This author justifies the 
	truncations of the Hubbard 
	model in the "atomic 
	limit". In this way, one 
	avoids the high complexity 
	(such as screening effects) 
	of the long-range Coulomb 
	forces.
\bibitem{all}
	D.B. McWhan, A. Menth, J.P. 
	Remeika, W.F. Brinkman, and 
	T.M. Rice, Phys. Rev. B {\bf 7}, 
	1920 (1973); S.A. Carter, 
	T.F. Rosenbaum, J.M. Honig, 
	and J. Spalek, Phys. Rev. Lett. 
	{\bf 67}, 3440 (1991); Y. 
	Tokura, Phys. Rev. Lett. {\bf 70}, 
	2126 (1993).
\bibitem{Alexei}
	Alexei M. Tsvelik, {\it Quantum Field 
	Theory in Condensed 
	Matter Physics}. Chapter 12. 
	Cambridge University Press (1995).
\bibitem{magnetic}
	T. Herrmann and W. Nolting, Phys. 
	Rev. B {\bf 53}, 10579 
	(1996). These authors discuss the 
	case of magnetism due to 
	quasi-particle damping which is 
	neglected in our work; a more recent 
	calculation by the same 
	authors is given 
	in Cond-Mat/9702022.
\bibitem{Nolting}
        W. Nolting,
        Z. Physik {\bf 225}, 
	25 (1972); W. Nolting,
        {\it Grundkurs: Theoretische
        Physik. 7 Viel-Teilchen-Theorie.} 
        Verlag Zimmermann-Neufang
        (Ulmen -1992).
\bibitem{Micnas-et-al}
	R. Micnas, M.H. Pedersen, S. 
	Schafroth, T. Schneider, 
	J.J. Rodr\'{\i}guez-N\'u\~nez 
	and H. Beck, Phys. Rev. 
	B {\bf 52}, 16223 (1995). 
	The authors use the moment 
	approach for the attractive 
	Hubbard model for 
	intermediate interaction. See also, 
	T. Schneider, M.H. Pedersen and 
	J.J. Rodr\'{\i}guez-N\'u\~nez 
	Z. Phys. B {\bf 100}, 263 (1996). 
	The last paper sheds some light 
	for $B = B(\vec{k})$, i.e., 
	beyond the spherical approximation 
	of Nolting\cite{Nolting}.
\bibitem{Izyumov-Sckryabin}
	Yu. Izyumov and 
	Yu. N. Skryabin, {\it Statistical 
	Mechanics of Magnetically 
	Ordered Systems}. New York: 
	Plenum Press 1988.
\bibitem{LRot}
	Laura Roth, Phys. Rev. 
	Lett. {\bf 184}, 451 (1969).
\bibitem{Scalapino}
	N. Bulut, D.J. Scalapino and 
	S.R. White, Phys. Rev. Lett. 
	{\bf 73}, 748 (1994). 
	They find that for $U/t = 8$, a	spin-density-wave 
	approximation provides a 
	sensible description of the spectral 
	weight, $A({\bf k},\omega)$, 
	at half filling. See also, S.R. 
	White, D.J. Scalapino, R.L. Sugar 
	and N.E. Bickers, Phys. Rev. 
	Lett. {\bf 63}, 1523 (1989) 
	where the authors use a new 
	method to perform the analytical 
	continuation from QMC data. They 
	get the two peak structure of a 
	SDW for lattices $4\times 4$ and 
	$8\times 8$ at $U/t = 4.0$, 
	$T/t = 1/12$ and $\rho = 1/2$. 
\bibitem{CM}
	Andrey V. Chubukov and Dirk K. Morr, 
	cond-mat/9701196.
\bibitem{KS}
	A.P. Kampf and J.R. Schrieffer, 
	Phys. Rev. B 
	{\bf 42}, 7967 (1990).
\bibitem{Bennemann}
	M. Langer, J. Schmalian, S. Grabowski 
	and K.-H. Bennemann, 
	Phys. Rev. Lett. {\bf 75}, 4508	(1995); 
	see also, S. Grabowski, 
	M. Langer, J. Schmalian and K.-H. 
	Bennemann, Europhys. Lett. {\bf 34}, 
	219 (1996).
\bibitem{Singer}
	J.M. Singer, M.H. Pedersen, 
	T. Schneider, 
	H. Beck and H.-G. 
	Matuttis, Phys. Rev. B {\bf 54}, 
	1286 (1996); see also, R. 
	Lacaze, A. Morel, B. Petersson and J. 
	Schr\"oper (Submitted to 
	J. Physique I (France)).
\bibitem{Elk}
	K. Elk and W. Gasser, 
	{\it Die Methode der 
	Greenschen Funktionen 
	in der Festk\"orphysik}. 
	Akademie - Verlag. 
	Berlin (1979). Section 8.4.
\bibitem{Vollhardt}
	Dieter Vollhardt, in {\it 
	Correlated Electron Systems}, 
	p. 57. World 
	Scientific (1993).
\bibitem{NJR}
	W. Nolting, S. Mathi 
	Jaya and S. Rex, 
	Phys. Rev. B {\bf 54}, 
	14455 (1996).
\end{references}
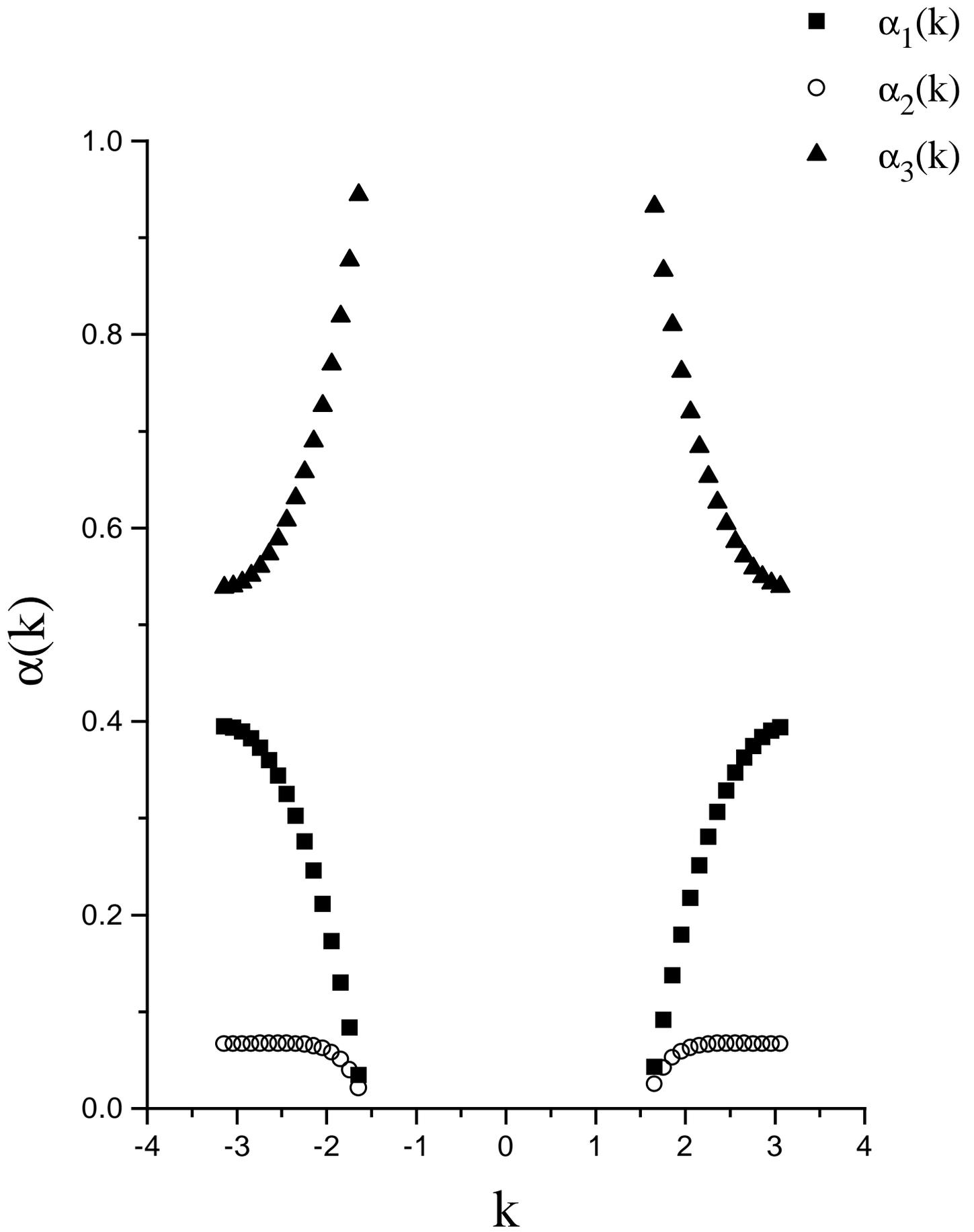
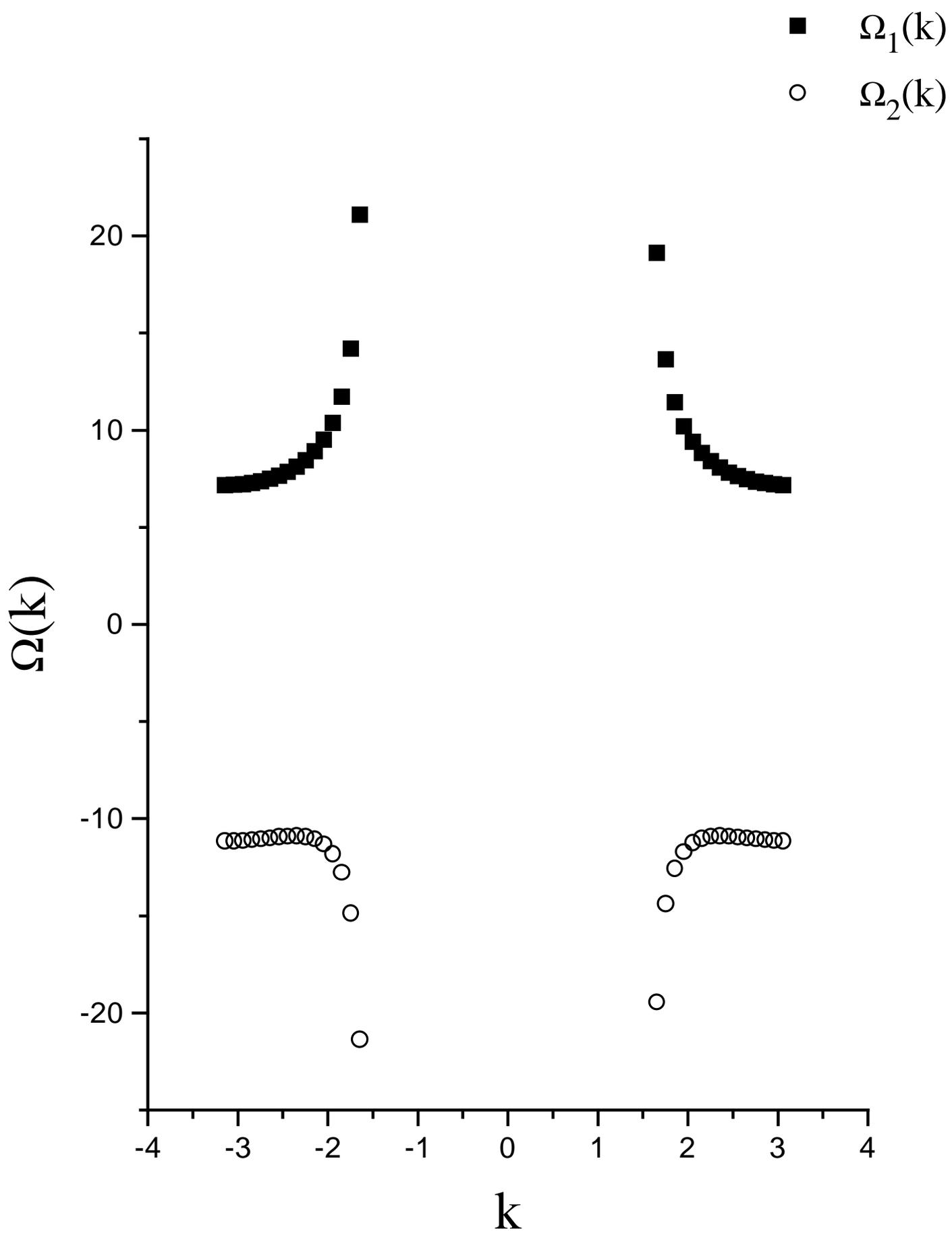
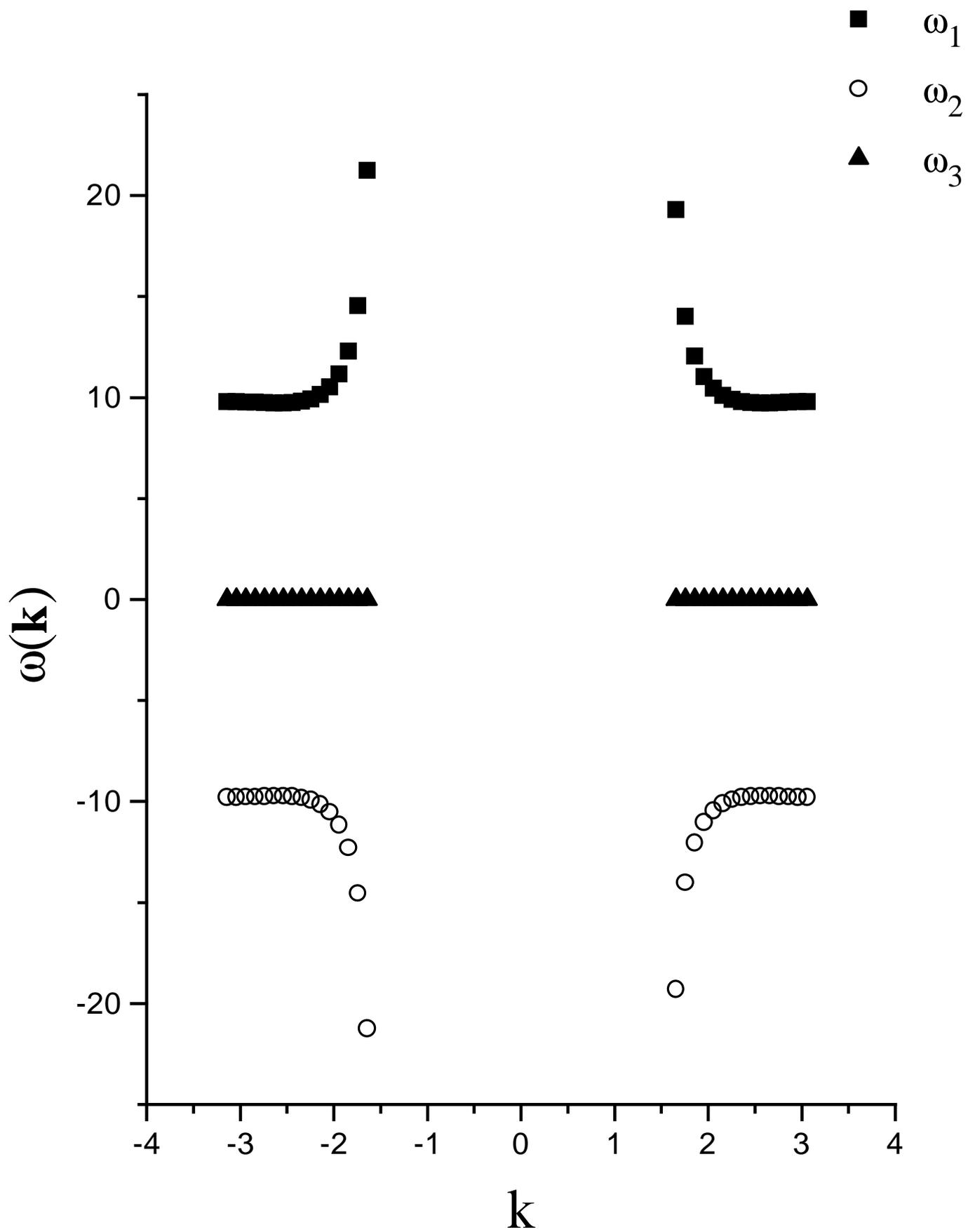
\end{document}